\documentclass{llncs}

\usepackage{cite}
\usepackage{amsmath,amssymb,amsfonts}
\usepackage{algorithm}
\usepackage{algorithmicx}
\usepackage{multirow}
\usepackage{algpseudocode}
\usepackage{subfig}
\usepackage{ifthen}
\usepackage{graphicx}
\usepackage{textcomp}
\usepackage{xcolor}
\usepackage{array}

\newcolumntype{P}[1]{>{\centering\arraybackslash}p{#1}}

\usepackage{hyperref}
\usepackage{epsfig}
\usepackage{psfrag}

\begin{document}

\title{\textsf{Split-n-Chain}: Privacy-Preserving Multi-Node \\ 
Split Learning with Blockchain-Based Auditability\thanks{This preprint has not undergone peer review (when applicable) or any post-submission improvements or corrections. The Version of Record of this article is published in Cluster Computing (Springer), and is available online at \href{https://doi.org/10.1007/s10586-026-06142-5}{https://doi.org/10.1007/s10586-026-06142-5}.}}

\author{
Mukesh Sahani\inst{1}
\and
Binanda Sengupta\inst{2}
}

\index{Sahani, Mukesh}
\index{Sengupta, Binanda}
\institute{
Indian Institute of Technology (Indian School of Mines) Dhanbad, India\\
\email{22mt0219@iitism.ac.in}
\and
Indian Institute of Engineering Science and Technology, Shibpur, India\\
\email{binanda@it.iiests.ac.in}
}

\maketitle

\begin{abstract}
Deep learning, when integrated with a large amount of training data, has the potential to outperform machine learning in terms of high accuracy. Recently, privacy-preserving deep learning has drawn significant attention of the research community. Different privacy notions, in the context of deep learning, include privacy of data provided by data owners and privacy of parameters and/or hyperparameters of the underlying deep neural network. Federated learning is a popular privacy-preserving execution environment where data owners participate in learning the parameters collectively without leaking their respective data to other participants. However, federated learning suffers from certain security/privacy issues: as each participant has access to the architecture of the whole neural network, a malicious participant can leak the architecture to a third party. In this paper, we propose a variant of split learning, that we call \textsf{Split-n-Chain}, where the layers (and respective parameters) of the neural network are split among several distributed nodes such that the whole network is not available to any single node. \textsf{Split-n-Chain} achieves several privacy properties: data owners need not share their training data with other nodes, and no individual node has access to the parameters and hyperparameters of the deep neural network (except those of the respective layer it holds). 
Moreover, \textsf{Split-n-Chain} exploits blockchain to audit the computation done by different (possibly malicious) nodes. 
We implement \textsf{Split-n-Chain} and experiment with the same in two settings: using a single local machine as well as a group of distributed machines connected through a LAN. Our experimental results show that: \textsf{Split-n-Chain} is efficient, in terms of time required to execute different phases, and the training loss trend in \textsf{Split-n-Chain} is similar to that for the same deep neural network when implemented in a monolithic fashion. 


\end{abstract}

\section{Introduction}
Recent developments in deep learning~\cite{DBLP:journals/neco/HintonOT06} and state-of-the-art 
processing units have led to unprecedented accuracy in various tasks including image classification and recognition \cite{DBLP:journals/pami/RenHG017,DBLP:conf/cvpr/GirshickDDM14}, drug discovery \cite{8982968} and speech recognition \cite{8793338}. Such a wide range of applications has resulted in the rise of a recent concept called 
``Deep Learning as a Service (DLaaS)''~\cite{WuLPCSWZ22}. 
DLaaS aims to provide deep-learning-based solutions and mechanisms with the help of cloud-based computing infrastructures. Deep neural networks are expected to be more accurate, in general, if trained with more diverse data --- which motivates companies and institutions to collect as much data as possible and feed it to these models. This data is usually generated by users' devices such as cell phones \cite{9172037}, wearable devices \cite{8995238} and pacemakers \cite{9662651}. However, this data may contain sensitive information that a user does not want to disclose to the model owner or other users. To train a deep neural network while protecting user privacy, researchers have come up with several solutions, such as differential privacy~\cite{9044259}, that have been applied to logistic regression \cite{9984869}, and $k$-mean clustering \cite{9799519}. In \cite{9833432}, homomorphic encryption techniques are used to train convolutional neural network (CNN) models privately over encrypted data. However, these techniques are computationally intensive and incur a high overhead in terms of training time.

In \cite{9478216}, Ray et al.~investigate 
collaborative deep learning with strong privacy protection while maintaining high data utility. Here, data owners participate in learning the parameters of a deep neural network collectively, where each participant benefits from the data provided by all participants. This technique of distributed and collaborative deep learning is known as ``federated learning''. Each participant has its data which is not sufficient to train a deep neural network with acceptable accuracy. In federated learning, many participants collaboratively train, with the help of a parameter-server, a model without sharing their respective data with others. The parameter-server trains a global model, and each participant has an identical local model at its end. Each participant trains its own local model using only the data it holds and sends an intermediate gradient to the parameter-server. Once the parameter-server receives intermediate gradients from all participants, it performs certain aggregation functions over the gradient values and computes global parameters. These parameters are again downloaded by each participant to be used in the next iteration. Federated learning suffers from certain privacy issues in that each participant knows the parameters and hyperparameters of the 
neural network.

Gupta and Raskar~\cite{DBLP:journals/jnca/GuptaR18} propose an alternative to federated learning called ``split learning'' in order to achieve privacy. Data-source nodes (clients) want to train a deep neural network over their data. The actual training is done over a server having large computational resources. ~\cite{DBLP:journals/jnca/GuptaR18} splits a neural network into two parts through a \textit{cut layer}. The first part of the training computation is performed on the client side and the second part is performed on the server side -- such that the server does not have access to the client's data, and the client as well as the server do not know all the parameters and hyperparameters of the neural network. However, both the client and the server still have access to significant portions of the parameters and hyperparameters of the network. 
Moreover,~\cite{DBLP:journals/jnca/GuptaR18} does not discuss how to select the cut layer appropriately in case a client is resource-constrained and may not have enough resources to perform expensive computations for multiple layers of the deep neural network.

One possible way to overcome the aforementioned issues is to split the deep neural network among multiple nodes (including one or more clients, the server, and certain additional intermediate nodes), where each node is given layers based on its computational capabilities. In the training process of the deep neural network, we make sure that the neural network encounters each data point in the training dataset \emph{epoch} number of times. We cannot feed all training data, possibly from multiple clients, into the neural network at the same time. So, we divide the training dataset into batches, and in each \emph{iteration} a batch of training data is fed into the neural network.
In case some of these nodes become malicious, that may degrade the accuracy of the neural network. Therefore, in addition to privacy, we require certain mechanisms, so that nodes' contribution during model training at different epochs and iterations  can be audited.

\medskip
\noindent
\textbf{Our contributions}: We summarize our major contributions towards designing a privacy-preserving multi-node split neural network as follows. 
\begin{itemize}
    \item We propose a privacy-preserving multi-node split-learning solution that we call \textsf{Split-n-Chain}, where a deep neural network is split among clients, a server and some intermediate nodes (unlike~\cite{DBLP:journals/jnca/GuptaR18}, where a network is split between a client and a server only). Each node in \textsf{Split-n-Chain} holds one or more layers of the neural network. Each of the clients holds some initial layers (including the input layer) of the network, and training data provided by one client is processed at a time. The server (or the terminal node) holds some final layers (including the output layer) of the network. All other intermediate layers are distributed among the intermediate nodes.
    \smallskip
    \item We propose an algorithm for the efficient distribution of layers among intermediate nodes. Intermediate layers of the neural network are assigned to intermediate nodes based on their computational capacity. During the initial bootstrapping phase of \textsf{Split-n-Chain}, multiple proof-of-work (PoW) problems are given to each intermediate node to quantify their computational power. \smallskip
    \item We introduce blockchain in the context of split learning to audit data contribution and layer computation done by different nodes. In \textsf{Split-n-Chain}, if we observe any malicious trend in the change of loss value between some specific iterations of an epoch, we can consult the blockchain to find which data-source nodes have contributed training data during that epoch. We design an incentive mechanism for intermediate nodes in the deep neural network. If the loss value of the network is reduced to a threshold value, then all intermediate nodes get incentives based on how many layers of the neural network they train. Thus, if an intermediate node acts maliciously, it reduces the accuracy of the neural network, and no intermediate nodes get the incentive. Here, the incentive mechanism acts as a driving force for intermediate nodes to behave honestly.\smallskip
    \item We implement \textsf{Split-n-Chain} in two different settings: on a local machine as well as over multiple personal computers connected through a local area network (LAN). Our experimental results show that \textsf{Split-n-Chain} is practical in terms of its performance. 
\end{itemize}

\medskip
The rest of the paper is organized as follows.
Section~\ref{sec:prelims} discusses certain preliminaries and background relevant to our work.
In Section~\ref{sec:scheme}, we provide a detailed construction of \textsf{Split-n-Chain}, our privacy-preserving multi-node split-learning scheme enabled with blockchain-based auditability.
Section~\ref{sec:performance_ana} discusses the performance of \textsf{Split-n-Chain} based on different parameters. 
In the concluding Section~\ref{sec:conclusion}, we summarize the work done in this paper.

\section{Preliminaries and Background}
\label{sec:prelims}
In this section, we give an overview of different aspects of deep learning and blockchain, 
relevant to our work. We enlist in Table~\ref{table:notation} different notations that we use in this paper. 
\renewcommand{\arraystretch}{1.2}
\begin{table}[htbp]
\centering
\caption{Notations used}
\begin{tabular}{|P{1.5cm}||P{10.0cm}|}
\hline
\textbf{Notation} & \textbf{Description} \\
\hline
$[x,y]$ & The set $\{x,x+1,\ldots,y\}$, where $x$ and $y$ are two integers with $x\le y$ \\
\hline
$l$ & Layer index of a deep neural network \\
\hline
$q_l$ & Number of neurons present at layer $l$ \\
\hline
$w_{i, j}^l$ & Weight of synaptic connection between $i$-th neuron of layer $l - 1$ and $j$-th neuron of layer $l$\\
\hline
$b_i^l$ & Bias value for $i$-th neuron of layer $l$ \\
\hline
$Y_{actual}$ & Observed value of the label-output of the neural network \\
\hline
$Y_{predicted}$ & Predicated value of the label-output of the neural network \\
\hline
$E_{training}$ & Error between $Y_{actual}$ and $Y_{predicted}$\\
\hline
$\eta$ & Learning rate of the neural network \\
\hline
$\frac{\partial E_{training}}{\partial w_{i, j}^l}$ & Partial derivative of $E_{training}$ with respect to $w_{i, j}^l$ \\
\hline
$p$ & Number of intermediate nodes considered for layer distribution\\
\hline
$node_j$ & $j$-th node in the computation chain \\
\hline 
$m$ & Number of proof-of-work (PoW) problems given to each intermediate node \\
\hline
$T_j^i$ & Time taken by $node_j$ to solve $i$-th PoW problem \\
\hline
$T_j$ & Total time taken by $node_j$ to solve all PoW problems given \\
\hline
$CP_j$ & Computational power of ${node_j}$ \\
\hline
$CPF_j$ & Computational power fraction of ${node_j}$ \\
\hline
$n$ & Total number of training samples \\
\hline
$TPC$ & Total parameters in the neural network\\
\hline
$NP_j$ & Number of parameters assigned to $node_j$ \\ 
\hline
$MLD$  & \textit{Model-layer description} is a list of layer descriptions of the model (each element of the list contains the description of a layer of the model) 
\\
\hline
$NLD$ & \textit{Node-layer description} is a list of elements where each element is a list containing the descriptions of the layers assigned to an intermediate node 
\\
\hline
$CPFList$ & List of computational power fractions of intermediate nodes (in descending order) \\
\hline
\end{tabular}
\label{table:notation}
\end{table}

\subsection{Deep Learning}
\label{subsec:dl}
Deep learning~\cite{DBLP:journals/neco/HintonOT06}, a subset of machine learning, performs representational learning --- which makes it reusable in various domains. A typical deep neural network consists of three types of layers: an input layer, hidden layers (possibly multiple), and an output layer. 
Based on the task we need to solve, there are different kinds of neural networks which include fully connected neural networks~\cite{DBLP:journals/tkde/MinnixMI92}, convolutional neural networks~\cite{DBLP:journals/pieee/LeCunBBH98}, sequential neural networks~\cite{DBLP:conf/icassp/Alvarez-Cercadillo93}.

\begin{figure}[t]
    \centering
    \includegraphics[width=5.7cm, height=5.0cm]{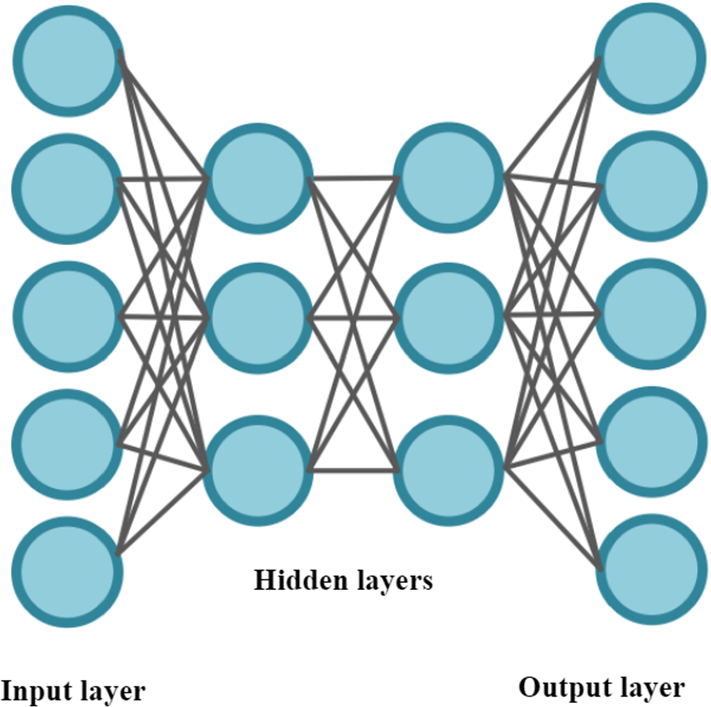}
    \caption{A fully connected neural network}
    \label{fig:neural_network}
\end{figure}

In a fully connected neural network, each layer $l$ contains a certain number (say, $q_l$) of neurons, and 
the output of neuron $i$ (for each $i\in[1,q_{l-1}]$) at layer $l-1$ connects to the input of each neuron at layer $l$ (see Figure~\ref{fig:neural_network}). For each such connection between two neurons, there is a weight assigned to it. For example, $w_{i,j}^l$ is a weight assigned to the connection between neuron $i$ at layer $l-1$ and neuron $j$ at layer $l$. Each neuron $j$ also has a bias $b_j^l$.

\textit{Backpropagation} \cite{10097504} is one of the most efficient learning methods for deep neural networks. Feedforward and backpropagation steps are executed, in loops, for a specific number of times. In the feedforward pass, the output of each layer is computed using the output of the previous layer and the parameters of the current layer. A key component of a deep learning neural network is the activation function which adds non-linearity to that network. So, neural networks can learn non-linear relationships between features and the label.

Backpropagation uses the gradient descent method, which attempts to reduce the error $E_{training}$ (this is a convex function -- such as mean squared error, root mean squared error -- that describes the difference between $Y_{predicted}$ and $Y_{actual}$). The error function is selected based on the type of task being performed. For example, we can consider the following \textit{mean squared error} (MSE) as an error function
$$E_{training}=\frac{1}{2}\sum\limits_{i=1}^N(Y_{predicted}  - Y_{actual})^2.$$  
Once we have $E_{training}$, we can use it for parameter optimization of the network. We can update $w_{i,j}^l$ as
$$w_{i,j}^l = w_{i, j}^l - \eta\frac{\partial E_{training}}{\partial w_{i,j}^l},$$ 
where $\eta$ is the learning rate and $\frac{\partial E_{training}}{\partial w_{i,j}^l} $ is the partial derivative of $E_{training}$ with respect to $w_{i,j}^l$. The learning process is repeated for a predefined number of times. The weights and biases associated with different layers are known as \textit{parameters} of the neural network, and information, such as the number of layers, number of neurons in each layer, activation functions, and the error function, are known as \textit{hyperparameters} of the network.

\subsection{Distributed Deep Learning}
\label{subsec:distdl}
Deep neural networks may have many (e.g., in the order of hundreds) layers, and training such a huge network is often a computationally intensive task. There exists a large literature on various distributed deep-learning techniques~\cite{8916416,8644603,9973543} which attempt to distribute the computational load over multiple nodes and train deep neural networks over a distributed environment. On the other hand, from the privacy point of view, federated learning \cite{9478216,9264412} and split learning \cite{DBLP:journals/jnca/GuptaR18} are two major distributed deep-learning techniques that we discuss in the following sections.

\subsubsection{Federated Learning}
\label{subsubsec:fl}
In federated learning~\cite{9478216,9264412}, a deep neural network is trained in a distributed fashion. Federated learning enables $N$ users, each having a certain amount of data, to train the network collaboratively without sharing their respective data with other users. The model-owner provides a copy of the model to each user. There is a centralized parameter-server that performs collaborative training. Each user trains its local model using its own data and sends intermediate gradient values to the parameter-server. Once the parameter-server receives intermediate gradients from all nodes, it performs certain aggregation methods to compute optimized weights of the deep-learning model. Each user downloads the updated weights from the parameter-server and updates its local model accordingly. This process is iterated for a specific (predefined) number of times. In federated learning, the parameters and hyperparameters of the underlying deep neural network is known to every user, and any attacker can get access to them by comprising a single user. Unlike federated learning, our goal in this work is to preserve privacy not only of users' individual data but also of these parameters and hyperparameters.

\subsubsection{Split Learning}
\label{subsubsec:sl}
In split learning~\cite{DBLP:journals/jnca/GuptaR18}, the layers of a deep neural network is split into two parts, partitioned using a cut layer. The layers before the cut layer are trained by the client and the rest of the layers are trained by the server. \cite{DBLP:journals/jnca/GuptaR18} describes split learning with multiple data sources and a single server, and discusses parameter synchronization between data sources for layers before the cut layer. During an epoch, one of the clients uses its data to compute the output of layers it holds and sends the output to the server. The server takes input from the client and uses this input to compute the output of the layers it possesses. After the output computation, the server computes the loss value, and uses the computed loss value to find the gradients required for weight optimization. It computes all the gradients till the cut layer and sends the gradient of the cut layer to the client. Given the gradient, the client performs weight optimization of all layers before the cut layer. This loop of operation is performed by each client for a specific number of times. 

We note that, even in split learning, the parameters and hyperparameters are exposed to the clients and the server. So, any attacker can get access to enough parameters and hyperparameters if any client is compromised. In this paper, we propose an approach to split the neural network further among multiple intermediate computing nodes which makes it harder for the attacker to compromise multiple nodes in order to learn similar information. We propose an efficient bootstrapping algorithm to split the layers of the neural network among the data-source, intermediate and terminal nodes. To audit these nodes for possible malicious activities, we employ blockchain in our solution (we give a brief overview of blockchain in Section~\ref{subsec:Blockchain}).

\subsection{Blockchain}
\label{subsec:Blockchain}
Blockchain~\cite{nakamoto2008bitcoin} has been a topic of interest for research communities as well as industries~\cite{8666530}. It is a decentralized, tamper-resistant, tamper-evident and time-ordered distributed ledger of blocks~\cite{8029379}. A block stores transactions which contain details of certain operations. Each block contains a timestamp and a reference to the previous block to maintain a chain-like structure among blocks. In Bitcoin \cite{nakamoto2008bitcoin}, a transaction contains information about money transfers (in terms of BTC) among pseudonymous users. Workers (or miners) collect transactions and compete with each other to produce a new block containing some of these transactions. The worker, who generates a valid block and appends it to the blockchain, gets a certain amount of reward (in terms of BTC). To decide the winning miner, proof of work (PoW)~\cite{9209934} is used. Proof of work is a consensus mechanism that requires a significant amount of effort (in terms of hash computation) from a network of nodes.

Apart from PoW, there exist several consensus mechanisms based on different principles such as proof of stake (PoS), Byzantine fault tolerance (BFT) and hybrid protocols. The important aspects of desiging a consensus protocol~\cite{9845078}, in the blockchain setting, include the following.
\begin{itemize}
    \item \emph{Leader selection}: Selection of the worker who will add the next block in the blockchain.
    \item \emph{Network model}: Synchronous, asynchronous, or semi-synchronous message communication.
    \item \emph{Communication complexity}: Cost involved in propagating a new block to all workers.
    \item \emph{System model}: Permissionless (any worker can join the network) or permissioned (only a worker with certain credentials can join the network) blockchain.
    \item \emph{Consensus property}: Agreement, validation, and authentication among nodes.
\end{itemize}

Algorand~\cite{10.1145/3132747.3132757,8987305} is a hybrid consensus protocol based on PoS and BFT, different from a PoW-based consensus protocol. Algorand guarantees consensus finality, i.e., a valid block appended to the chain is never removed in the future --- which is suitable especially for our problem. Without removing any block from the blockchain, we can avoid spending time and computational resources to retrain the deep-learning model. In the next section, we discuss the construction of \textsf{Split-n-Chain} which integrates blockchain with split learning in order to achieve auditability. We also use PoW to estimate the computational power of a node in order to compute the number of layers to be assigned to that node.

\section{\textsf{Split-n-Chain}: Our Proposed Solution}
\label{sec:scheme}
In this work, we propose a new approach to train a deep neural network over multiple data-source nodes, multiple intermediate nodes (where each intermediate node holds one or more layers of the neural network) and a terminal node (owned by the DLaaS owner) while mitigating the need to share raw labeled data directly (we refer to Section~\ref{subsec:SystemOverview} for detailed description of these nodes). Moreover, no data-source nodes or intermediate nodes do not have access to the hyperparameters of the neural network. On the other hand, for auditability purposes, each node's layer computation must be registered as a transaction on the underlying blockchain used in \textsf{Split-n-Chain}. We aim to provide a solution while satisfying the following requirements.
\begin{enumerate}
    \item A single data-source node does not need to share the raw labeled data with any other data-source node, intermediate node, or terminal node.
    \item The DLaaS owner controls the architecture of the neural network.
    \item Each intermediate node is provided with one or more layers of the network.
    \item  The data-source node controls the parameters of initial layers, and the terminal node controls the parameters final layers of the neural network.
    \item  No data-source node or intermediate node should have access to the whole architecture of the neural network; only the 
    DLaaS owner has access and control over the architecture of the neural network.
    \item No data-source node, intermediate node, or terminal node should have access to all the trained parameters of the neural network. 
    \item  Any computation performed by the data-source node, intermediate node, and terminal node over the deep neural network layers should be registered as one or more transactions on the blockchain.
\end{enumerate}
In the following section, we describe \textsf{Split-n-Chain}, our scheme to split and train a deep neural network over the data-source, intermediate, and terminal nodes. Our techniques also include methods to quantify the computing power of each intermediate node, encode data into different spaces, and transmit them to train the neural network. Our algorithm can be run using one or more data- source nodes, one or more intermediate nodes, and a single terminal node.

\subsection{System Overview}
\label{subsec:SystemOverview}
\textsf{Split-n-Chain} combines split learning, cryptographic primitives and blockchain techniques, in a novel way, to achieve secure, distributed and privacy-preserving deep learning. Suppose there are $N$ data-source nodes (clients) who want to train a deep-learning model owned by a DLaas 
owner in a privacy-preserving way, but any client individually does not have enough data. So, they need to train the deep-learning model collaboratively without revealing their data to any other data-source nodes or the model owner. If every data-source node has access to the model and trains a collaborative model using a setting like \emph{federated learning}, then it has access to all the parameters and hyperparameters. In that case, even if a single data source is compromised, the parameters and hyperparameters of the model will be leaked. \emph{Split learning} is proposed to overcome this issue with federated learning in the following way. The neural network is split into two parts: initial layers are controlled by a data source (or client), and the rest of the layers are controlled by a server owned by the DLaaS owner. However, in this setting also, data-source nodes hold significant amount of parameters/hyperparameters of the network.

\begin{figure}[t]
    \centering
    \includegraphics[width=\textwidth, height=0.35\textheight]{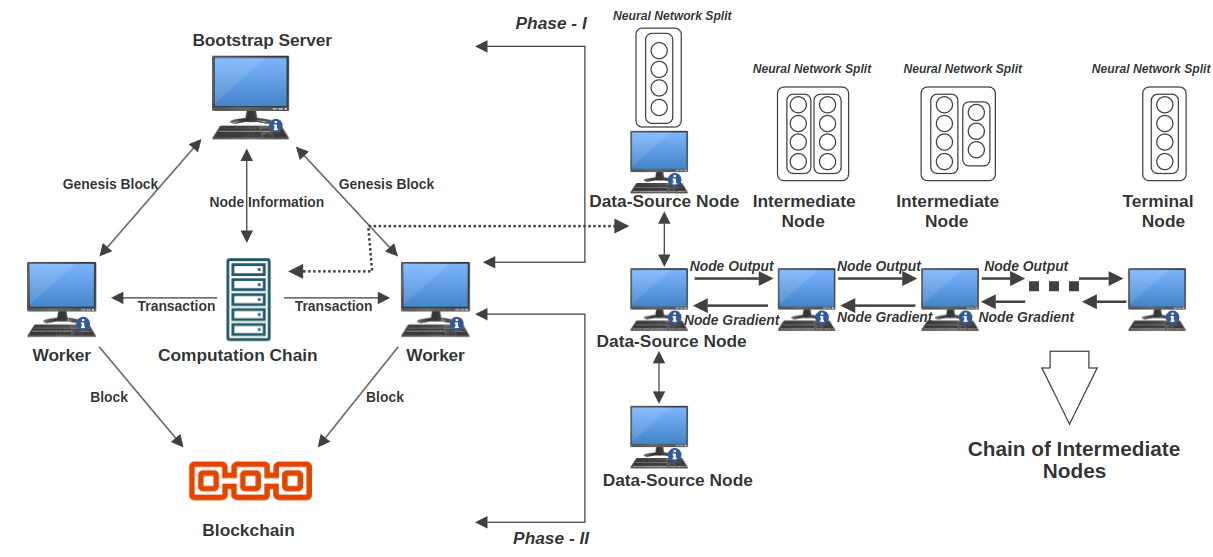}
    \caption{System overview of \textsf{Split-n-Chain} }
    \label{fig:overview-label}
\end{figure}

We propose \textsf{Split-n-Chain} where the layers of the neural network is split among data-source nodes, intermediate nodes, and terminal nodes. 
A data-source node contain some initial layers and the terminal node (owned by the DLaaS owner) contain some final layers. The rest of the layers are split among the intermediate nodes. 
We first describe different entities involved in \textsf{Split-n-Chain}  
(see Figure~\ref{fig:overview-label}) 
and then discuss some of its design principles. 
Finally, we describe, in details, our solution \textsf{Split-n-Chain} and different phases it involves.
\begin{itemize}
    \item \textbf{Data-source node}: Data-source nodes (or clients) are those entities which own the data required to train the given deep neural network. 
    There can be one or more data-source nodes involved.
    In \textsf{Split-n-Chain}, some of the initial input layers are assigned to these nodes --- which avoids passing the raw data to the next node in the computation chain. The data provided by all data-source nodes is used for training the network, with one such node processed at a time.
    \item \textbf{Intermediate node}: 
    Intermediate nodes are provided with intermediate layers of the neural network in order to perform training computation for those layers.
    They can either be personal computers or high-performance servers. Depending on the system requirements, there can be zero or more intermediate nodes involved. 
    \item \textbf{Terminal node}: The terminal node (or server), owned by the DLaaS owner, contains some of the final layers of the neural network. This node computes the final output and initiates the optimization of the network along the computation chain, but in the reverse order.
    \item \textbf{Bootstrap server}: A bootstrap server is responsible for distributing the layers of deep neural network among data-source nodes, intermediate nodes and terminal nodes. It has the information about the hyperparameters of the deep neural network. 
    \item \textbf{Worker node}: Worker nodes (or simply, workers) are responsible for validation and verification of transactions created by the data-source, intermediate and terminal nodes while they compute the outputs and gradients of their respective layers. Workers compete among themselves in order to add a block of transactions to the underlying blockchain. The worker, who finishes the job first, becomes a leader. The leader then adds its block to the blockchain.
    \item \textbf{Computation chain}: This is a chain of nodes that performs the sequential computation in order to train the deep neural network. That is, the output of a node in the chain is provided to the next node as its input (and the gradient values are passed in the reverse order). Some initial layers (including the input layer) are assigned to each data-source node and some final layers (including the output layer) are assigned to the terminal node. Hidden (intermediate) layers are distributed among the intermediate nodes. This specific order of the nodes in the computation chain is important for the correctness of the computation during both forward and backward passes.
\end{itemize}

\textsf{Split-n-Chain} enables \textit{privacy-preserving} training of the neural network in the sense that: (1) no data-source node needs to share its data with the model owner or any intermediate node or any other data-source node, (2) no data-source (or intermediate) node knows about the configuration of the whole computation chain and the hyperparameters of the neural network, and (3) no data-source node (or intermediate) node has access to all parameters of the neural network. 
After each layer computation and weight optimization, the corresponding node creates a transaction and sends it to a neighbor worker node. The worker validates the transaction, and the transaction is added to the blockchain upon successful validation. If the DLaaS owner detects any malicious behavior during an epoch of training, it uses that specific transaction stored in the blockchain for auditing purpose. This design principle of \textsf{Split-n-Chain} must address the following questions.
\begin{enumerate}
    \item How to get the data-source nodes registered for training the deep-learning model?
    \item How to inform the data-source, intermediate and the terminal nodes about the respective layers of the model they have to train?
    \item How to inform nodes about their neighbors (i.e., the \textit{next} nodes they have to send the output of their layers to and receive gradients from, and the \textit{previous} nodes they receive input for their layers from and send gradients to for weight optimization)?
    \item How to enable each node to create a dynamic neural network using the hyperparameters of its layers?
\end{enumerate}

The following sections describe the execution environment of \textsf{Split-n-Chain} which is split into two phases. \textit{Phase I} consists of bootstrapping which performs the initialization of the computation chain involving the data-source, intermediate nodes and terminal nodes (see Section~\ref{subsec:phaseI}). 
In \textit{Phase II}, the actual training of the neural network takes place where we train the deep neural network over the computation chain and each node creates transactions for auditing (see Section~\ref{subsec:phaseII}).

\subsection{Phase I: Bootstrapping}
\label{subsec:phaseI}
Boostrapping consists of four stages: \textit{collection of node information}, \textit{estimation of computational power of intermediate nodes}, \textit{splitting neural network} and \textit{generation of the genesis block}. All these steps are performed by a bootstrap server which is aware of the hyperparameters of the deep neural network. All data-source nodes, intermediate nodes, and the terminal node connect to the bootstrap server to register themselves on the computation chain over which the neural network will be trained. The bootstrap server has information regarding the configuration of the neural network which includes the details of the layers controlled by the data-source nodes, the terminal node, and layers to be distributed among the intermediate nodes (see Table \ref{table:model_config}). 
\begin{table}[tbp]
\centering
\caption{Model configuration}
\begin{tabular}{|P{1.9cm}||P{10.0cm}|}
 \hline
  \textbf{Attribute}& \textbf{Value} \\
\hline
Model name & Basic split-learning model\\
\hline
Input shape &  (3, 128, 128)\\
\hline
Layer description (data-source node) & [\{``LayerType'': ``Conv2D",
        ``LayerDesc": \{``filter": 32,
            ``kernel": 5,
            ``padding": 2, 
            ``dropout": 0.4,
            ``activation": ``ReLU"\}]\\
\hline
Layer description (intermediate node) &  [\{``LayerType": ``Conv2D",
            ``filter": 64,
            ``kernel": 5,
            ``padding": 2,
            ``dropout": 0.5,
            ``activation": ``ReLU"\},
        \{``LayerType": ``Con2D",
            ``filter": 64,
            ``kernel": 5,
            ``padding": 2,
            ``dropout": 0.5,
            ``activation": ``ReLU" \},
        \{``LayerType": ``Conv2D",
            ``filter": 128,
            ``kernel": 5,
            ``padding": 2,
            ``dropout": 0.5,
            ``activation": ``ReLU"\},
        \{``LayerType": ``Conv2D",
            ``filter": 128,
            ``kernel": 5,
            ``padding": 2,
            ``dropout": 0.5,
            ``activation": ``ReLU"\},
        \{``LayerType": ``Flatten"\}]\\
\hline
Layer description (terminal node) & [\{``LayerType": ``FullyConnected",
            ``InputShape":784,
            ``OutputShape":128,
            ``Activation": ``ReLU"\},
        \{``LayerType": ``FullyConnected",
            ``InputShape": 128,
            ``OutputShape": 1,
            ``activation": ``SigMoid"\}]\\
\hline
\end{tabular}
\label{table:model_config}
\end{table}
Each of the four stages is discussed as follows.

\begin{itemize}
    \item \textbf{Collection of node information}: In this stage, each node, that wants to be a part of the computation chain or wants to be a worker for processing transactions generated by nodes, sends the following attributes to the bootstrap server.
    \begin{itemize}
        \item \emph{Node port number}: It denotes the port number at which the node will listen to the previous node in the computation chain. For worker nodes, it is the port number at which it will listen for transactions.
        \item \emph{Node type}: Three types of nodes are involved in the computation chain: data-source nodes, intermediate nodes, terminal nodes, and worker nodes involved in processing transactions. A node declares the type of node it wants to be.
        \item \emph{Node public key}: The public key of a node is used for authenticating the transactions created by that node during layer-output computation and weight optimization. 
        \item \emph{Number of data points}: If the type of a node is ``data-source'', then it should declare the number of data points it holds. 
\end{itemize}

At this point, the bootstrap server possesses all the information required to create a computation chain which requires the layers to be distributed among nodes. It is straightforward for the bootstrap server to distribute equal numbers of layers to the intermediate nodes and initialize the computation chain. However, this approach is suitable mostly for a homogeneously distributed environment where all intermediate nodes have comparable computational resources; whereas in practice, we may have nodes with different computational power. Thus, to cope with this requirement, we need to quantify the relative computational power of these nodes and distribute layers to nodes based on their respective (relative) computational power. We estimate the computational power of intermediate nodes in the next stage discussed below. 
\medskip

\item \textbf{Estimation of the computational power of intermediate nodes}: In this stage, we can ask intermediate nodes to provide an estimation of their computational power in terms of floating point operations per second (FLOPS). However, a malicious node can claim to have more computational power than the actual to get more layers (and thus more incentive). 
Then, that specific node can become a bottleneck in the computation chain. Instead, in \textsf{Split-n-Chain}, the bootstrap server gives every intermediate node a certain number of (computational) problems to solve. To be precise, the bootstrap server gives $m$ problem to each node and, based on how much time a node takes to solve those problems, we can estimate the relative computational of the intermediate nodes as follows. The bootstrap server computes the total time taken by node$_j$ to compute the solution of all problems given to the intermediate node as
\begin{align}
    T_j = \sum_{i=1}^m T_j^i.
\end{align}
As we have 
\begin{align}
    CP_{j} \propto \frac{1}{T_j}, 
\end{align}
the bootstrap server uses this computational power of node$_j$ (i.e., $CP_{j}$) to estimate the fraction of computational power of node$_j$ (i.e., $CPF_j$), with respect to other intermediate nodes, as 
\begin{align}
\label{equation:comp_frac}
    CPF_j = \frac{CP_{j}}{\sum_{i=1}^{p}CP_{i}}, 
\end{align}
where $p$ intermediate nodes are considered for layer distribution. Based on this estimation of the computation power fraction of each intermediate node, the bootstrap server splits the neural network among intermediate nodes in the next stage discussed below. 
\medskip

\item \textbf{Splitting neural network}: 
In this stage, the bootstrap server distributes the hidden (intermediate) layers among intermediate nodes, so that the number of parameters assigned to $node_j$
is directly proportional to its computational power fraction, i.e., 
\begin{gather}
\label{equation:node_param_count}
    NP_j\approx TPC*CPF_{j}.
\end{gather}
The bootstrap server assigns to $node_j$ a certain number of layers, such that the number of parameters present in those layers become greater than or equal to $TPC*CPF_j$. 
This is because \textsf{Split-n-Chain} does not split the parameters of a single layer among multiple intermediate nodes. Thus, any intermediate node should get that many layers that make its parameter count at least the quantity shown in Eqn.~\ref{equation:node_param_count}.

On the other hand, the number of intermediate nodes may be more than that of the intermediate layers. We sort intermediate nodes based on their computational power fraction to resolve such conflicts. If there are more intermediate nodes than the number of intermediate layers, priority is given to a node with a higher 
$CPF$ value 
(see Eqn.~\ref{equation:comp_frac}).

Algorithm~\ref{alg:cap} shows an algorithm to distribute the intermediate layers of a fully connected neural network among intermediate nodes, based on their computational power fractions. 
\begin{algorithm*}[htbp]
\caption{An algorithm to distribute hidden layers of a fully connected neural network among intermediate nodes 
}
\label{alg:cap}
\begin{algorithmic}
\Require $MLD$: Model-layer description (layer descriptions are shown in Table~\ref{table:model_config})
\Require $CPFList$: Computational power fraction list of intermediate nodes (in descending order)
\smallskip
\State $NLD \leftarrow []$  //Comment: $NLD$ contains layer description for each intermediate node; this list is empty initially 
\State $TPC \leftarrow 0$ //Comment: $TPC$ contains the total number of parameters in the deep neural network \\
\\
//Comment: Find the total number of parameters in the deep neural network
\For{$i \leftarrow 1$ to $MLD.length$} 
    \State $TPC = TPC + MLD[i - 1].neuron\_count * MLD[i].neuron\_count $
    \EndFor
\\
\State $i \leftarrow 1$
\For{$k \leftarrow 1$ to $CPFList.length$} //Comment: For each intermediate node, find the number of layers assigned to it
    \State $NP_k \leftarrow 0$
    \If{$k == CPFList.length$} //Comment: If only one intermediate node is left, assign all remaining layers to it
        \State $NLD[k] = MLD[i : ]$
    \Else
        \State $NP_k$ = $TPC * CPFList[k]$ //Comment: Compute the number of parameters to be assigned to a node
        \State $nodeparamcount \leftarrow 0$
        \State {$j \leftarrow i + 1$}  //Comment: Select layers until the number of parameters becomes equal or greater than $NP_k$ 
        \While{$nodeparamcount \leq NP_k$ and $j < MLD.length$} 
            \State $nodeparamcount = nodeparamcount + MLD[j - 1].neuron\_count * MLD[j].neuron\_count$
            \State {$j \leftarrow j + 1$}
        \EndWhile
        \State $NLD[k] = MLD[i: j])$ //Comment: Add layer description to $NLD$
        \State{$i \leftarrow j$} //Comment: Move to the last layer that is not assigned to an intermediate node yet
    \EndIf
\EndFor
\State \Return {$NLD$}
\end{algorithmic}
\end{algorithm*}
$MLD$ (\textit{model-layer description}) is a list of layer descriptions of the model where each element of the list contains the description of a layer of the deep neural network. $MLD[i]$, $MLD[i:j]$ and $MLD[i:]$ contain the description of the $i$-th layer, the descriptions of the $i$-th layer to the $j$-th layer, and the descriptions of all layers starting from the $i$-th layer, respectively.
$MLD$ contains descriptions of layers 
as shown in Table~\ref{table:model_config}. For simplicity, in Algorithm~\ref{alg:cap}, {\textcolor{black}{we assume that
$MLD[i]$ contains hyperparameters for the $i$-th layer, such as the number of neurons present in the layer, information on the activation function, and dropout applied after layer computation.}} 
\textit{CPFList} contains the computational power fractions, in descending order, of all intermediate nodes. 
$NLD$ (\textit{node-layer description}) is a list of elements where each element is a list containing the descriptions of the layers assigned to an intermediate node. $NLD[i]$ is a list containing the description of layers assigned to the $i$-th intermediate node in the computation chain.
$TPC$ (total parameter count) is the total number of trainable parameters in the deep neural network. $NP_k$ is the number of parameters to be allocated to $node_k$.

After processing Algorithm~\ref{alg:cap}, the bootstrap server has the necessary information to split the network. The bootstrap server creates a computation chain and sends each node the following information.
\begin{itemize}
    \item \emph{Next-node IP address and port number}: This information consists of the IP address and port number of the next node in the computation chain to which a node has to send its layer computation output and take gradient as input to perform layer weight optimization. 
    Details of the next node's IP address and port number help a node to connect its next node in the computation chain. 
    \item \emph{Node-layer description}: %List of layers description, 
   This information contains \emph{layer type} and other layer attributes based on the layer type (e.g., for a fully connected neural network, it contains \emph{number of neurons in the layer} and \emph{activation function}; for convolutional neural network, it contains \emph{number of filters}, \emph{kernel shape}, \emph{padding}, \emph{activation function}). 
    Using the layer description list, a node becomes aware of the layers it needs for training.
\end{itemize}

\medskip

\item \textbf{Generation of the genesis block}: After the bootstrap server creates the computation chain, information of this chain is registered as a block in the blockchain (this block serves as the \emph{genesis block} of the blockchain). 
The bootstrap server includes the following information in the genesis block.
\begin{itemize}
    \item \emph{Model name}: This is the name given to a specific deep neural network (e.g., MNIST Digit Classifier).
    \item \emph{List of addresses of data-source nodes}: 
    The address of a data-source node is the SHA256 hash of the public key of the data-source node.
    \item \emph{List of intermediate node}: This is a list of addresses of the intermediate nodes (it need not be in the same order as in the computation chain). Here, the address of an intermediate node is the SHA256 hash of the public key of the intermediate node.
    \item \emph{Address of terminal node}: The address of the terminal node is the SHA256 hash of the public key of the terminal node.
\end{itemize}
After the genesis block is ready, the bootstrap server digitally signs the genesis block and broadcasts the signed block to the workers of the blockchain. Workers then validate and authenticate the genesis block. If found valid, the blockchain is initialized by the genesis block. 
The bootstrap server then initializes the computational chain. All data-source, intermediate, and terminal nodes now know the layers they need to train and the next node in the computation chain (except the terminal node as it is the last node in the computation chain). We discuss the training phase of \textsf{Split-n-Chain} in Section~\ref{subsec:phaseII}.
\end{itemize}

\subsection{Phase-II: Training}
\label{subsec:phaseII}
Once bootstrapping is done, each node knows the hyperparameters of layers assigned to them and also the information of neighbors in the computation chain. Phase-II of \textsf{Split-n-Chain} trains the 
neural network split over the computation chain. 
Phase-II is divided into two stages: \emph{building the split neural network}, \emph{training the split neural network}. 
\begin{itemize}
    \item \textbf{Building the split neural network}: Each node builds a (partial) neural network using the information of the layers assigned to it and initializes the weights of these layers randomly (we call it a split neural network), so that it can perform forward and backward passes over this split neural network. 
    \medskip
    \item \textbf{Training the split neural network}: After all nodes in the computation chain are ready with their split neural networks, training is performed in a way similar to training a monolithic neural network (but with the neural network distributed among multiple nodes). The neural network is trained for a specific number of epochs decided by the bootstrap server. For each epoch, every data-source node sends its data for training in a batch of sizes already specified by the bootstrap server. In each epoch and for every batch of data from the data-source node, \textsf{Split-n-Chain} performs \emph{forward passes} and a \emph{backward passes}.
    \begin{itemize}
        \item \emph{Forward pass}: A forward pass starts with a data-source node. The data-source node takes a batch of data and finds the output of the split neural network it holds. After the output is produced the data-source node creates a blockchain transaction containing information shown in Table~\ref{table:output-transaction-info-table}. 
        \begin{table}[tbp]
        \centering
        \caption{Output-transaction information table}
        \begin{tabular}{|P{3.0cm}||P{8.0cm}|}
        \hline
        \textbf{Attribute}& \textbf{Value} \\
        \hline
        Node-type create transaction & {Data-source/intermediate/terminal node}\\
        \hline
        Epoch & Training epoch number \\
        \hline
        Batch number & Batch number of data of data-source node\\
        \hline
        Layer-output hash & SHA256 hash of output of the layers\\
        \hline
        Node address & SHA256 hash of public key of node\\
        \hline
        \end{tabular}
        \label{table:output-transaction-info-table}
        \end{table}
It signs the transaction and broadcasts it to all the workers in the blockchain. The data-source node then transmits layer output and label for training to the next node in the computation chain. The next node receives input from the previous node, uses this input to find the output of the split neural network it holds, and sends the output to the next node in the computation chain. Then, it creates (and broadcasts) a signed transaction as done by the data-source node. This process continues until the terminal node receives input from its previous node. It uses this input to find the output of the split neural network it holds and computes the predicted value for all data points in that batch. It also creates and broadcasts a transaction. The terminal node computes a loss value using a loss function and uses this loss value to compute the gradient needed for the optimization of the layers it holds. Once the computation of the loss value and gradient is done, the terminal node starts backward pass (backpropagation).\medskip

        \item \emph{Backward pass}: After the terminal node computes the gradient, it performs optimization of its layers using the gradient. It then creates (and signs) a transaction containing information shown in Table~\ref{table:gradient-transaction-info-table} and broadcasts the transaction to all the workers in the blockchain. 
        \begin{table}[tbp]
        \centering
        \caption{Gradient-transaction information table}
        \begin{tabular}{|P{3.0cm}||P{8.0cm}|}
        \hline
        \textbf{Attribute}& \textbf{Value} \\
        \hline
        Node-type create transaction & {Data-source/intermediate/terminal node}\\
        \hline
        Epoch & Training epoch number \\
        \hline
        Batch number & Batch number of data of data-source node\\
        \hline
        Node-gradient hash & SHA256 hash of gradient till initial layers of split neural network\\
        \hline
        Node address & SHA256 hash of public key of node\\
        \hline
        \end{tabular}
        \label{table:gradient-transaction-info-table}
        \end{table}
The terminal node computes the gradient till the initial layer of split neural network it holds and sends the gradient to the previous node in the computation chain. The previous node takes the gradient value, performs weight optimization, and finds the gradient till the initial layer of the split neural network it holds. It creates and signs a transaction similar to that done by the terminal node, and broadcasts it to all the workers in the blockchain. It also sends the gradient to the previous node. This continues until a gradient is received by the data-source node. Using this gradient value the data-source node performs weight optimization and finds the gradient till the initial layer of the split neural network it holds. It then creates and signs transaction and broadcasts them to all the workers in the blockchain.

\end{itemize}
 Forward and backward passes are performed \emph{epoch} numbers of times for each data point from each of the data-source nodes. Once we have a trained \textsf{Split-n-Chain}, we can use it to perform inference over the model. The DLaaS owner can perform validation of the model using some validation data. 
\end{itemize}

 \noindent
 \textbf{Incentive mechanism:} In \textsf{Split-n-Chain}, the money inflow in the network is done by the DLaaS owner which later provides the trained network, as a service, to its customers. Monetary incentive is the driving force to make every intermediate node in \textsf{Split-n-Chain} perform the honest computation of layers 
 assigned to it. If the loss value of training data at the terminal node converges below a predefined threshold value $\tau$, then each intermediate node in the computation chain gets an incentive directly proportional to the number of parameters it holds. If the training error loss is higher than $\tau$, then no intermediate nodes get any incentive. A worker node gets an incentive when it adds a new block to the blockchain. \textsf{Split-n-Chain} uses the consensus algorithm of Algorand
 (see Section~\ref{subsec:Blockchain}) which gives every worker node a fair chance to earn an incentive.

\section{Performance Evaluation}
\label{sec:performance_ana}
In this section, we provide a detailed description of \textsf{Split-n-Chain} implementation. We first discuss our implementation methodology and then report the experimental results. We evaluate the performance of \textsf{Split-n-Chain} in terms of various parameters such as bootstrapping time, training time and accuracy.

\subsection{Implementation Methodology}
\label{subsec:implementation_methodology}
We describe the implementation of \textsf{Split-n-Chain} in the following three modules: \textsf{BootstrappingAlgorithm}, \textsf{SplitLearningTrainingAlgorithm} and \textsf{SplitLearningBlockchainSystem}.

First, we build the \textsf{BootstrappingAlgorithm} module. The code for the bootstrap server is written in Python (version 3.8.10). It is implemented as a TCP server that waits for a specific number (this number may be provided as a command line argument as well) of TCP connection requests from a data-source node, intermediate node, terminal node, and worker node. Then bootstrap server collects node connection information from all participating nodes (including worker nodes) and stores this information in a dictionary where the IP addresses of nodes act as keys. Then, the bootstrap server sends every intermediate node a certain number ($m$) of proof-of-work (PoW) problems to solve (similar to the PoW problems used for mining in Bitcoin's blockchain~\cite{nakamoto2008bitcoin}). In \textsf{Split-n-Chain}, the bootstrap server provides intermediate nodes with random \emph{problem strings} of predefined length and difficulty level. The intermediate nodes solve all PoW problems given to them and send back the solution to the bootstrap server. After receiving all PoW solutions from each intermediate node, the bootstrap server splits the neural network among intermediate nodes according to Algorithm \ref{alg:cap} and sends the required information to these nodes. So, these nodes can now become part of the computation chain. After this, the bootstrap server generates a genesis block and broadcasts this genesis block to all workers of the blockchain. We note that, after this dissemination of genesis block, the bootstrap server is not required further in the following steps.

Second, we build \textsf{SplitLearningTrainingAlgorithm} module using  Pytorch~\cite{PyTorch_Documentation}.
We define the following 
primitive operations: \emph{receiveinput}, \emph{sendoutput}, \emph{receivegradient}, \emph{sendgradient} and \emph{createtransaction}. 
Each node, except a data-source node, first performs \emph{receiveinput} in order to receive input from its previous node in the computation chain. Then, the node performs layer computation to get an output and uses \emph{sendoutput} operation to send output to the next node (except the terminal node). After that, the node waits until it gets a gradient value, using \emph{receivegradient} operation, from its next node  in the chain. It then performs weight optimization and sends a gradient value to the previous node in the computation chain using \emph{sendgradient}. Before performing \emph{sendgradient} and \emph{sendoutput} operations, a node creates a transaction using \emph{createtransaction} operation and broadcasts this transaction to all workers of blockchain over UDP connections. All nodes in the computation chain perform almost the same operations except the data-source and terminal nodes. A data-source node computes the output of the initial layers using its own data, sends the output to the next node in the computation chain using \emph{sendoutput} operation, and waits until it receives a gradient from the next node using \emph{receivegradient} operation. After receiving the gradient, the data-source node performs weight optimization and starts the loop again, if required. On the other hand, the terminal node receives input from the previous node using \emph{receiveinput} and performs layer computation to predict the output. After finding the output, terminal node computes the loss value and finds the gradient for its layers. It then performs weight optimization and starts the backpropagation. The terminal node uses \emph{sendgradient} to send a gradient value to the previous node. These operations required to train the neural network are performed for a specific number of times depending on the number of epochs and iterations.

Third, we build \textsf{SplitLearningBlockchainSystem} where workers perform transaction validation and authentication. Nodes in the computation chain send transactions to workers over UDP connections using \emph{createtransaction} operation. Workers receive these transactions, validate them and check their authenticity using digital signatures attached to these transactions. We use the Algorand consensus algorithm~\cite{10.1145/3132747.3132757} to select a leader who is eligible to add a block to the blockchain. 
The selected leader then generates a block, containing transactions, to be added to the blockchain.

\subsection{Experimental Results}
We perform our experiments for \textsf{Split-n-Chain} in two different settings. We first implement it in a local system and show the results. Then, we experiment with multiple personal computers connected over a local area network (LAN). For training the neural network, we have used the standard Modified National Institute of Standards and Technology (MNIST) dataset~\cite{lecun-mnisthandwrittendigit-2010}, a dataset of handwritten digits. In the following sections, we discuss our experimental setup and results for both local and LAN setups.

\subsubsection{Implementation on a Local Machine}
\label{subsubsec:Imple_Local_Machine}
We simulate \textsf{Split-n-Chain} on a local machine where we dedicate a process to each individual node (i.e., each of the data-source, intermediate, terminal, worker, and bootstrap nodes). Processes communicate with each other over network protocols described in Section~\ref{subsec:implementation_methodology}. We use the MNIST dataset: 60000 samples are used for training, and 10000 samples are used for validation. We use a fully connected neural network for MNIST digit classification. 
Table~\ref{table:neural_network_split} shows a split of a fully connected neural network, that contains 55818 parameters, among a data-source node, two intermediate nodes and a terminal node. 
\begin{table}[tbp]
\centering
\caption{Splitting neural network with two intermediate nodes}
\begin{tabular}{||P{0.2\linewidth}||P{0.3\linewidth}||P{0.3\linewidth}||}
\hline
\textbf{Intermediate Node ID} &  \textbf{Computation Power Fraction} & \textbf{Number of Parameters} \\
\hline
\centering 1 & 0.30588235294 & 2080 \\
\hline
\centering 2 &  0.6941176470 & 3168 \\
\hline
\end{tabular}
\label{table:neural_network_split}
\end{table}
In this split, the data-source node contains 50140 parameters, the intermediate node 1 (that has a computational power fraction of 0.30588235294) contains 2080 parameters, the intermediate node 2 (that has a computational power fraction of 0.6941176470) contains 3168 parameters, and the terminal node contains 430 parameters. For the MNIST dataset, the data-source node controls a high number of parameters because of the high dimensional nature of image data.

\begin{figure*}[tbp]
      \centering
      \subfloat[Training loss without neural network split]{\includegraphics[width=.39\textwidth]{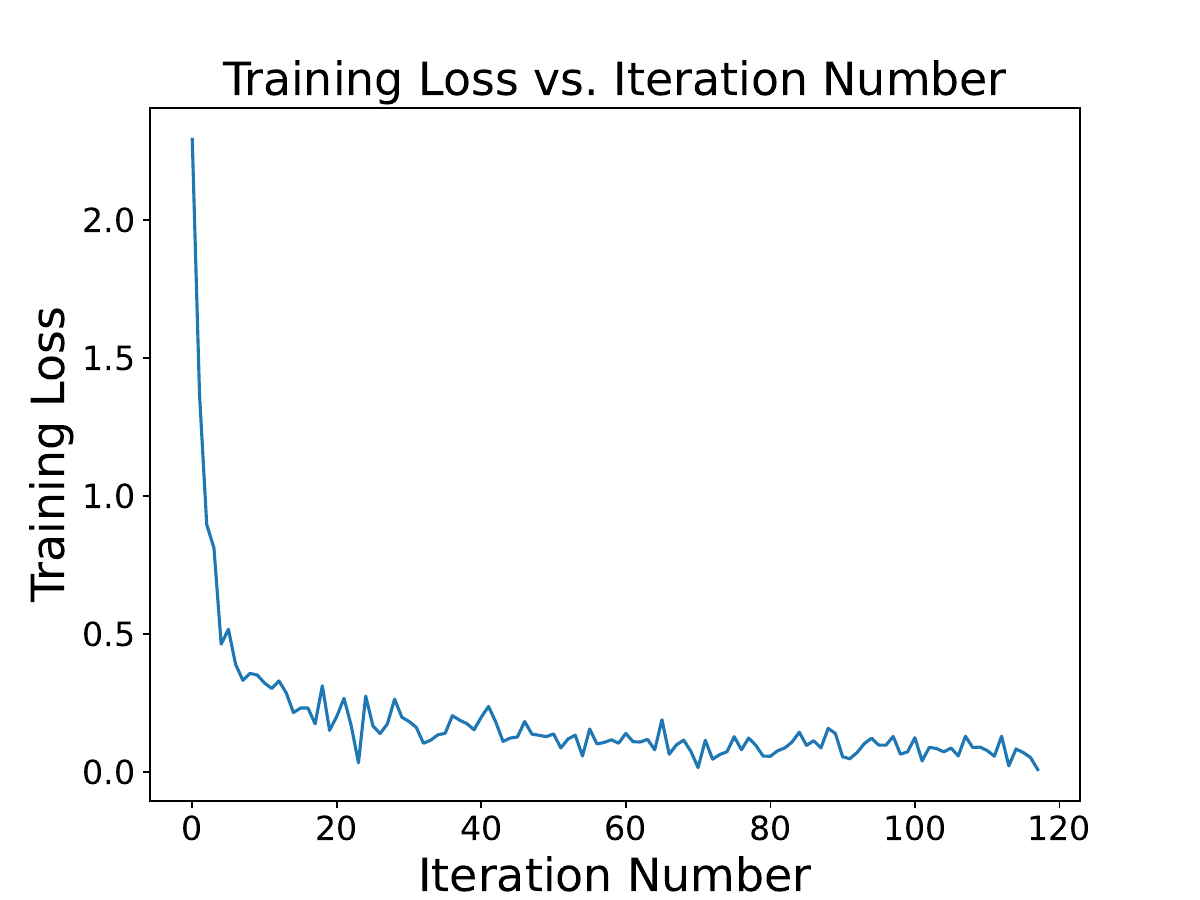}}
      \qquad
      \subfloat[Training loss with a split using two intermediate nodes]{\includegraphics[width=.39\textwidth]{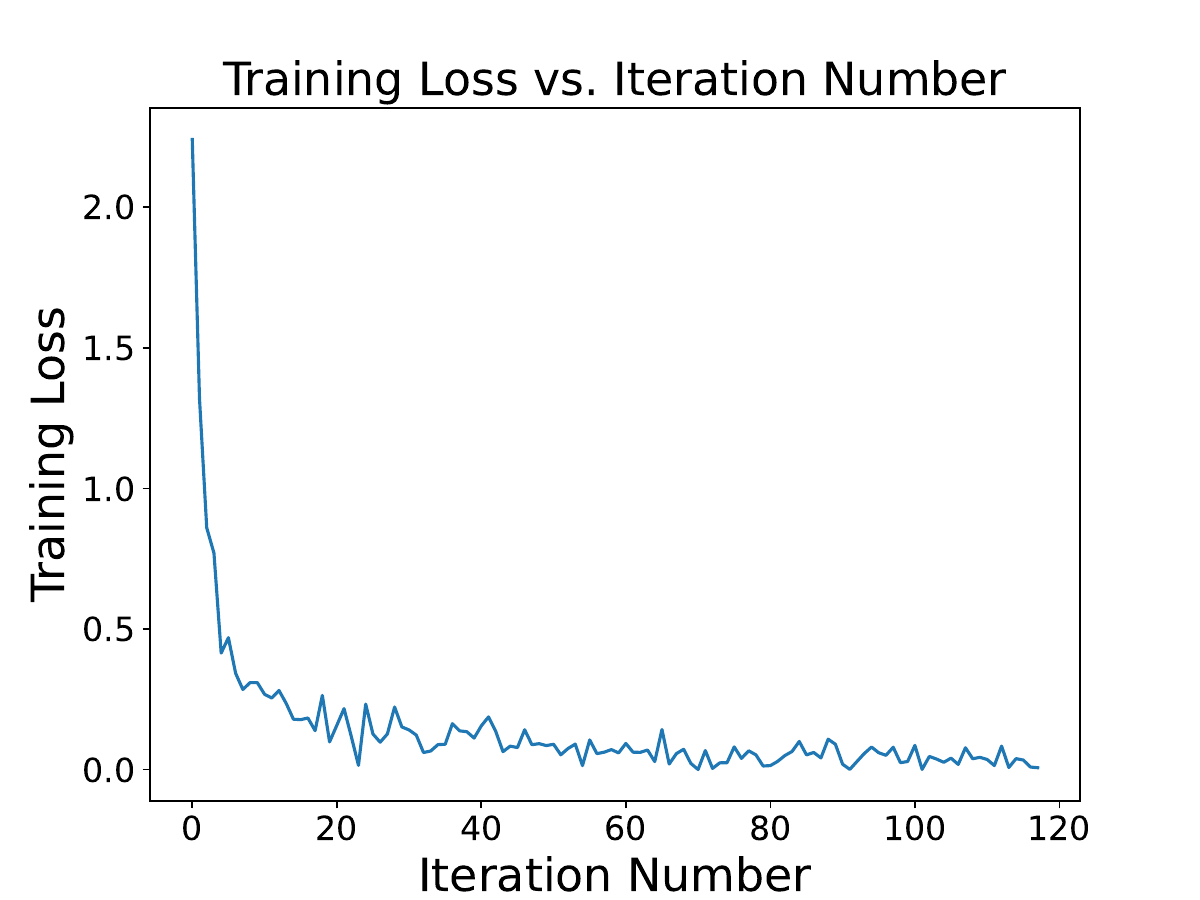}}
      \qquad
      \subfloat[Training loss with a split using three intermediate nodes]{\includegraphics[width=.39\textwidth]{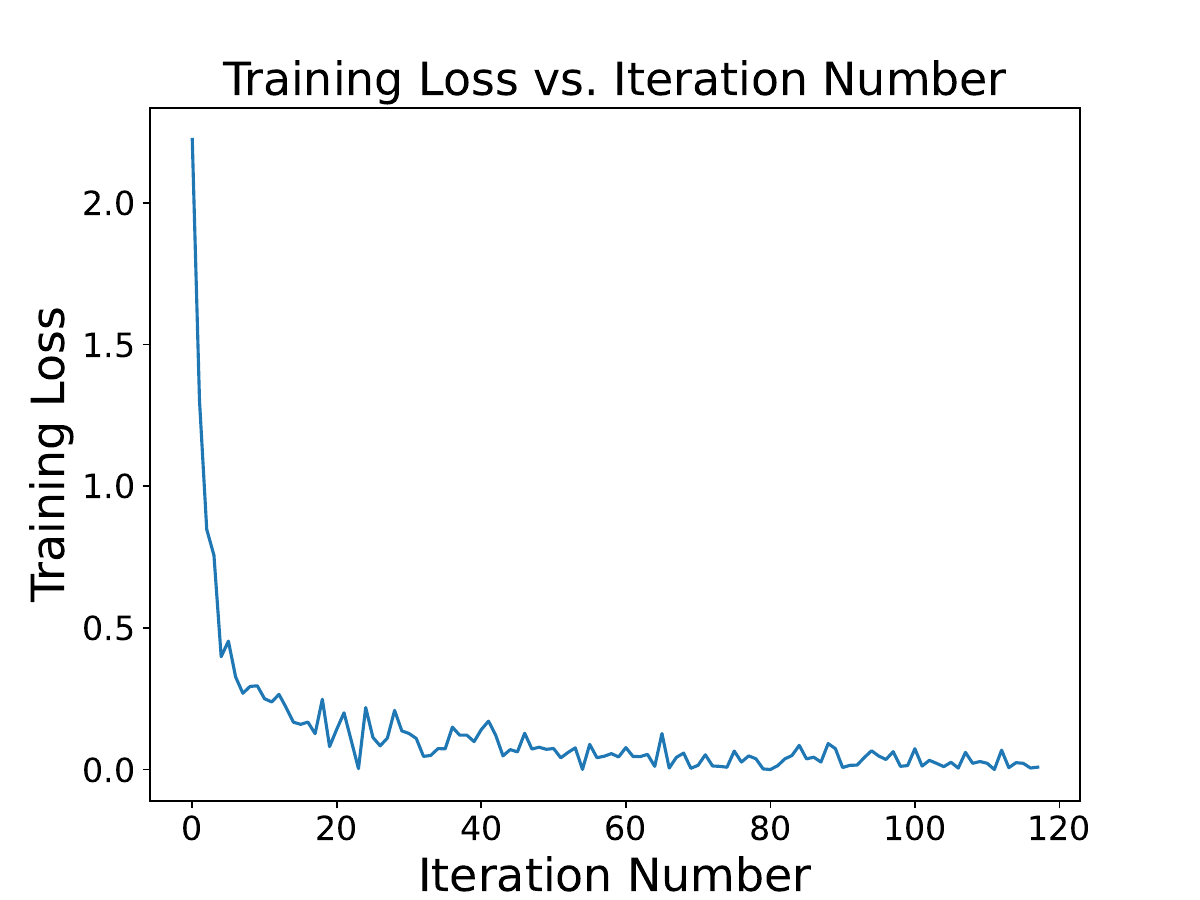}}
  \caption{Training loss vs.~iteration number when simulated on a local system}
\label{fig:training_loss_local_system}
\end{figure*}

We divide training into batches of size 512 and we train the neural network for 5 epochs. Each epoch consists of 118 iterations (since there are a total of 60000 training samples and each iteration takes 512 samples, each epoch needs 118 iterations in order to consider all training samples). Figure~\ref{fig:training_loss_local_system} shows the training loss trend in different scenarios: (1) without any split, (2) split using two intermediate nodes and (3) split using three intermediate nodes. 
\setlength{\tabcolsep}{3pt}
\begin{table}[tbp]
\caption{Different loss values and time for varying number of intermediate nodes (over a local machine)}
\centering
\begin{tabular}{||P{0.30\linewidth}||P{0.125\linewidth}||P{0.15\linewidth}||P{0.14\linewidth}||P{0.14\linewidth}||}
\hline
\textbf{Number of Intermediate Nodes} & \textbf{Training Loss} & \textbf{Validation Loss} & \textbf{Training Time (seconds)} & \textbf{Bootstrap Time (seconds)}\\
\hline
\centering 2 & \centering 0.124242 & \centering 0.127818 & \centering 54  & 127  \\
\hline
\centering 3 & \centering 0.115528 & \centering 0.121736 & \centering 64  & 141   \\
\hline
\end{tabular}
\label{table:local_machine_result}
\end{table} 

The training loss function trend is almost the same for these scenarios. Thus, there is no impact of network split on training loss and accuracy. Table \ref{table:local_machine_result} shows the training loss, validation loss, training time and bootstrapping time for \textsf{Split-n-Chain} with two and three intermediate nodes.
We observe that the split does not affect the training and validation loss. However, the training time for the neural network with three intermediate nodes is more than that with two intermediate nodes. This is due to the network delay introduced by one additional intermediate node. The bootstrapping time for the neural network with three nodes is more than that with two nodes due to the time spent by the additional node to solve its PoW problems.

\subsubsection{Implementation over a LAN}
\label{subsubsec:Imple_LAN}
We implement \textsf{Split-n-Chain} using multiple personal computers (PCs) connected over a LAN. We dedicate a PC to each individual node (i.e., each of the data-source, intermediate, terminal, worker and bootstrap nodes). We use the MNIST dataset with a split same as discussed in Section~\ref{subsubsec:Imple_Local_Machine} (i.e., 60000 samples for training and 10000 samples for validation). We experiment with the same samples and neural networks used in the local-system setting. 

\begin{figure*}[th]
      \centering
      \subfloat[Training loss with a split using two intermediate nodes]{\includegraphics[width=.35\textwidth]{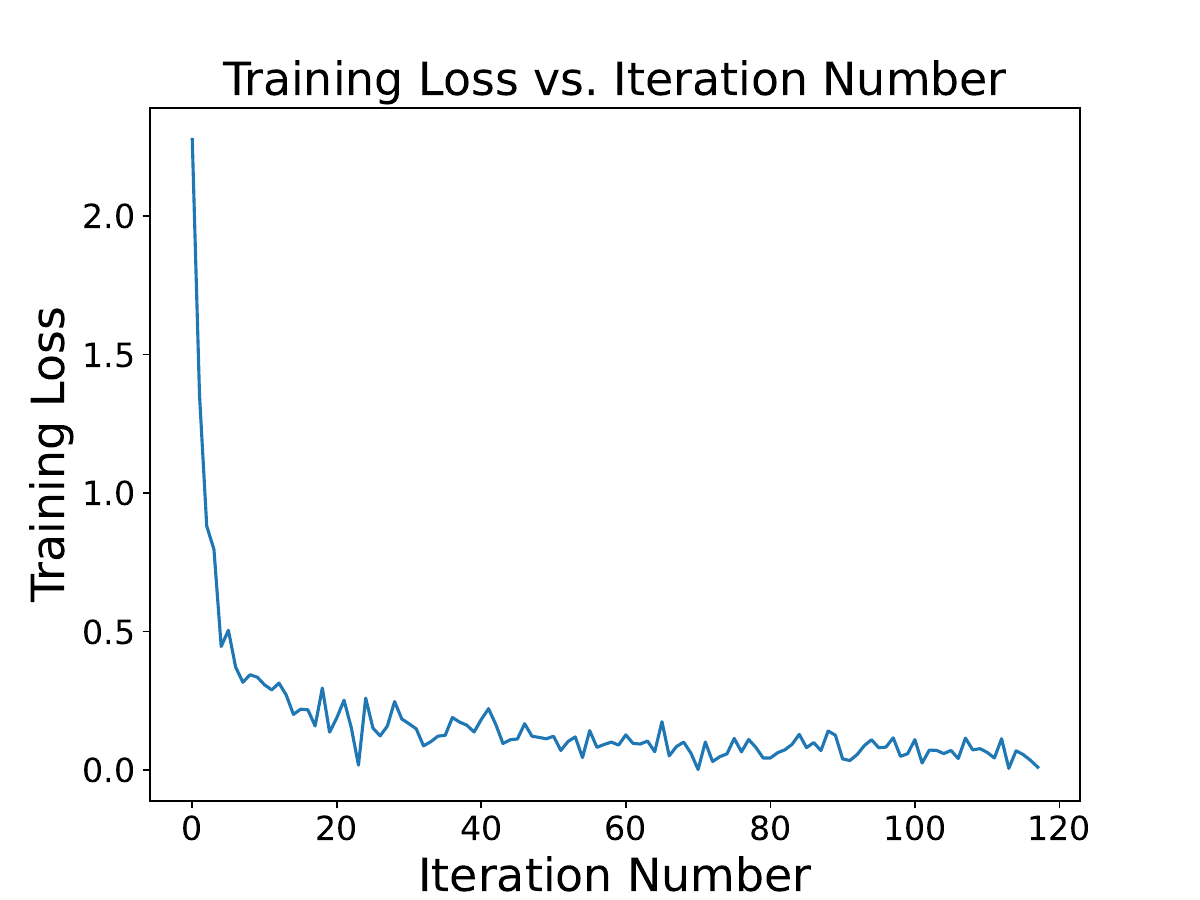}}
      \qquad
      \subfloat[Training loss with a split using three intermediate nodes]{\includegraphics[width=.35\textwidth]{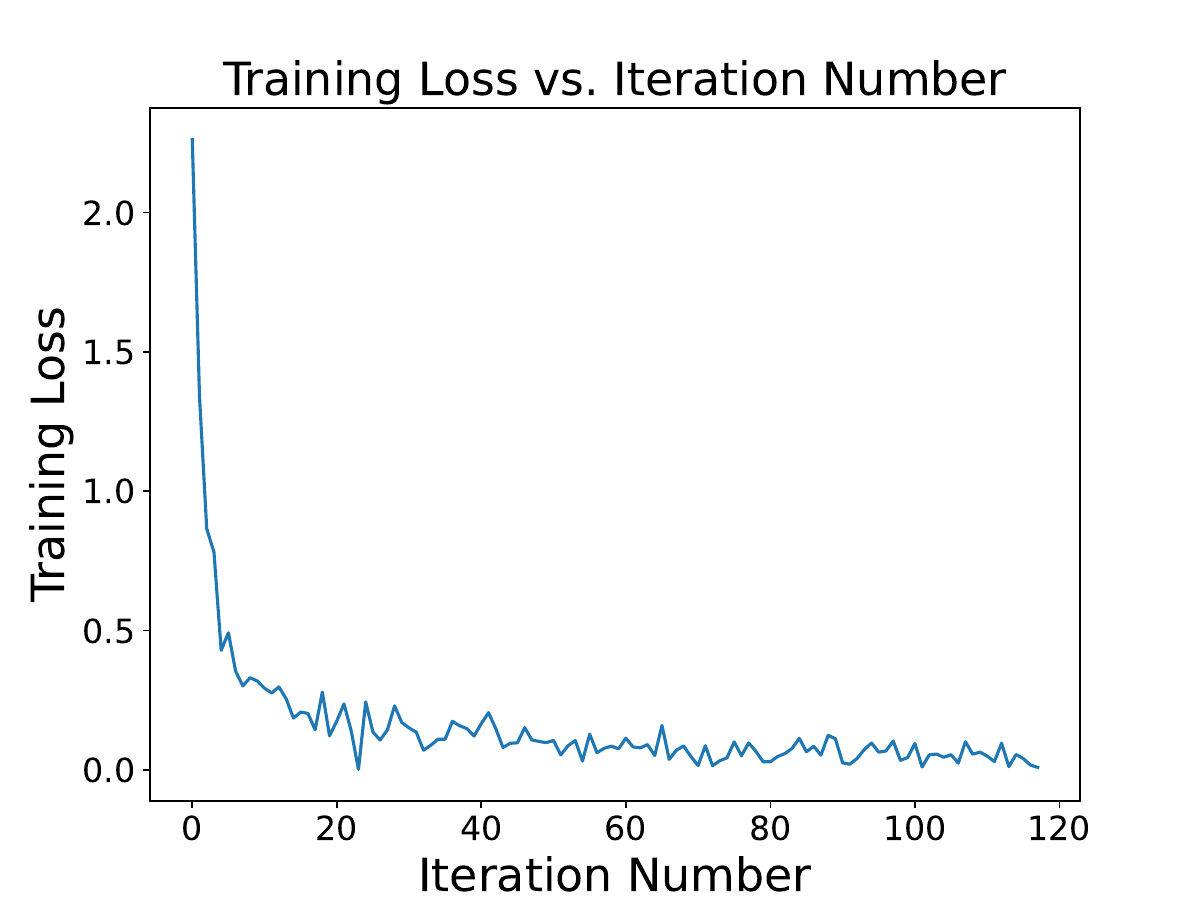}}
  \caption{Training loss vs.~iteration number when simulated over a LAN}
  \label{fig:training_loss_lan_network}
\end{figure*}

We report, in Table~\ref{table:lan_network_result}, the results obtained when we train the neural network over a LAN with two and three intermediate nodes, respectively. 
Figure~\ref{fig:training_loss_lan_network} shows the respective training loss trends. 
%with two and three intermediate nodes. 
It shows that the training loss trend is similar to that in the neural network without any split (see Figure~\ref{fig:training_loss_local_system}(a)). 
However, compared to the simulation on a local machine, both the training time and the bootstrapping time increase due to the additional transmission latency among the PCs connected through the LAN (see  Table~\ref{table:local_machine_result} and Table~\ref{table:lan_network_result}).

\begin{table}[tbp]
\caption{Different loss values and time for varying number of intermediate nodes (over a LAN)}
\centering
\begin{tabular}{||P{0.30\linewidth}||P{0.125\linewidth}||P{0.15\linewidth}||P{0.14\linewidth}||P{0.14\linewidth}||}
\hline
\textbf{Number of Intermediate Nodes} & \textbf{Training Loss} & \textbf{Validation Loss} & \textbf{Training Time (seconds)} & \textbf{Bootstrap Time (seconds)}\\
\hline
\centering 2 & \centering 0.136015 & \centering 0.138749 & \centering 74  & 640  \\
\hline
\centering 3 & \centering 0.116112 & \centering 0.131685 & \centering 97  & 770  \\
\hline
\end{tabular}
\label{table:lan_network_result}
\end{table}

\section{Conclusion}
\label{sec:conclusion}
In this paper, we design \textsf{Split-n-Chain}, a privacy-preserving multi-node split-learning solution that splits a deep neural network among multiple data-source nodes, intermediate nodes and a terminal node. 
\textsf{Split-n-Chain} employs an algorithm to distribute the intermediate layers of the neural network among the intermediate nodes present in the computation chain based on their computational efficiency (which is measured by, given certain PoW problems, how fast these nodes solve them). 
\textsf{Split-n-Chain} provides several privacy guarantees pertaining to the data owned by respective data-source nodes, the configuration of the whole computation chain and the parameters/hyperparameters of the neural network. 

For experimental validation of \textsf{Split-n-Chain}, we implement it in two different environments: on a local machine as well as on multiple PCs connected through a LAN. In the latter case, we get the training and validation loss values similar to those obtained for a local machine  --- which shows that splitting the neural network using multiple intermediate nodes does not affect the correctness of the network. Our experimental results validate that our solution can be used in practice to design a deep neural network enabled with multiple privacy guarantees.

%\bibliographystyle{splncs03} 
%\bibliography{draft}

\end{document}